\documentstyle[12pt]{article}

\begin{document}

\begin{flushright}
IMSc/2007/02/1 
\end{flushright} 

\vspace{2mm}

\vspace{2ex}

\begin{center}
{\large \bf Entropy of Anisotropic Universe } \\ 

\vspace{2ex}

{\large \bf and Fractional Branes } \\

\vspace{8ex}

{\large  S. Kalyana Rama}

\vspace{3ex}

Institute of Mathematical Sciences, C. I. T. Campus, 

Tharamani, CHENNAI 600 113, India. 

\vspace{1ex}

email: krama@imsc.res.in \\ 

\end{center}

\vspace{6ex}

\centerline{ABSTRACT}
\begin{quote} 

We obtain the entropy of a homogeneous anisotropic universe
applicable, by assumption, to the fractional branes in the
universe in the model of Chowdhury and Mathur. The entropy for
the 3 or 4 charge fractional branes thus obtained is not of the
expected form $E^{\frac{3}{2}}$ or $E^2$. One way the expected
form is realised is if $p \to \rho$ for the transverse
directions and if the compact directions remain constant in
size. These conditions are likely to be enforced by brane decay
and annihilation, and by the S, T, U dualities. T duality is
also likely to exclude high entropic cases, found in the
examples, which arise due to the compact space contracting to
zero size. Then the 4 charge fractional branes may indeed
provide a detailed realisation of the maximum entropic principle
we proposed recently to determine the number $(3 + 1)$ of large
spacetime dimensions.

\end{quote}

\vspace{2ex}


\newpage

\vspace{4ex}

{\bf 1. Introduction}  

\vspace{2ex}

We assume that our observed universe is described by string/M
theory. At low temperatures, the evolution of the universe is
described in the standard way by a low energy effective action.
At early times, when the temperature is of the order of string
scale, higher modes of the strings are excited and the evolution
must be described by stringy variables \cite{bowick,k1}.
Assuming that our universe originated from highly excited and
highly interacting strings, we have recently proposed in
\cite{k2} a maximum entropic principle to determine the number
$(3 + 1)$ of large spacetime dimensions. (For earlier ideas on
this number, see \cite{bv}.)

The evolution of the universe at early times, when its
temperature is high, is not well understood. Within the context
of perturbative string theory, a natural idea has been to assume
that the universe consists of gas of strings, their winding
modes, and/or gases of various branes, and then obtain the
evolution of the universe using a low energy effective action
\cite{bv, branes, b}.

Recently, Chowdhury and Mathur proposed in \cite{cs} a novel
model where the early universe consists of mutually BPS
intersecting brane and antibrane configurations. The branes (and
similarly antibranes) in these configurations form bound states
and become fractional, supporting very low energy excitations
and creating thereby large entropy for a given energy $E$. The
entropy $S$ of an $N$ charge fractional brane configuration is
expected to be \footnote{ This relation is derived with the
volume of the system kept fixed. However, for $N = 3, 4$
configurations considered here, it turns out to be valid in an
expanding universe also as explained later.} $\simeq
E^\frac{N}{2}$. Hence, the early universe is likely to contain,
and be dominated by, such fractional brane configurations
because of their high entropy. Chowdhury and Mathur also obtain
the energy momentum tensor, $T^\mu \; _\nu = diag \; (- \rho, \;
p_i)$ where $p_i = w_i \rho$, for these configurations, and
solve in complete generality the Einstein's equations of motion
with arbitrary $w_i$'s. See \cite{b} also.

The 4 charge fractional branes, with entropy $S$ expected to be
$\simeq E^2$, may provide a detailed realisation of the maximum
entropic principle proposed in \cite{k2} and may help in
understanding the evolution of the universe at early times. In
this paper we, therefore, study the relation between the entropy
$S$ and the energy $E$. We assume that the evolution of the
fractional branes in the universe is given by the evolution of
the anisotropic universe with the corresponding $w_i$'s; and
that the physical quantities $S$ and $E$ are similarly related.

Following the standard method, we obtain the entropy $S$ of the
anisotropic universe in general, as well as in the asymptotic
limit of large $t$. We consider two examples, particular cases
of which correspond to the fractional branes in the universe
with their $w_i$s obtained as in \cite{cs}. Using the results
of \cite{cs}, we then compare the entropy $S$ with its expected
form $\simeq E^\frac{N}{2}$ for the cases $N = 3, \; 4$. We find
that they do not agree. 

Given the importance of 4 charge fractional branes in
understanding the early universe, it is important to study how
to obtain the expected form for $S$. Hence, we consider a few
possible ways of obtaining the expected form for $S$. One of
them is that it is the holographic entropy which should obey the
expected relation.  Another is the following. Our examples show
that the expected form can be realised if $w_i = \omega \to 1$
for the transverse directions and if the compact directions
remain constant in size. We discuss the physical interpretation
of these conditions and argue, in the light of our two examples,
that they are likely to be enforced by including brane decay and
annihilation processes and by the S, T, U duality symmetries of
the string and M theories.

Also, there are high entropic cases in our examples, for example
$S \simeq E^X$ with $X > 4$, which arise due to the compact
space contracting to zero size. The T duality symmetry of string
theory is likely to prevent such contractions. We show that the
entropy then remains in accord with one's expectation that the
most entropic object is a Schwarzschild black hole or an
isotropic universe containing matter with $\omega = 1$.

This paper is organised as follows. In sections {\bf 2} and {\bf
3}, we present briefly the relevant equations of motion for an
anisotropic universe and their general and asymptotic solutions.
In section {\bf 4}, we present the expressions for entropy and,
in section {\bf 5}, two examples. In section {\bf 6}, we discuss
in detail the fractional branes. In section {\bf 7}, we conclude
by mentioning a few issues for further study.

\vspace{4ex}

{\bf 2. Equations of motion}  

\vspace{2ex}

Consider a $D$ -- dimensional homogeneous anisotropic universe
containing matter with $T^\mu \; _\nu = diag \; (- \rho, \;
p_i)$ where $\rho > 0$, $p_i = w_i \rho$ with $w_i$'s in the
range $- 1 \le w_i \le 1$, and $i = 1, 2, \cdots, D - 1$. The
line element is given by
\begin{equation}\label{ds}
d s^2 = - d t^2 + \sum e^{2 \lambda_i (t)} d x_i^2 
\end{equation}
where, here and in the following, the sum is over $i = 1, 2,
\cdots, D - 1$ unless mentioned otherwise. Defining $\Lambda =
\sum \lambda_i$ and $b = \sum w_i \lambda_i$ and using natural
units with $8 \pi G = 1$, the conservation equation $\nabla_\mu
T^\mu \; _\nu = 0$ and the Einstein's equations of motion
$R_{\mu \nu} - \frac{1} {2} g_{\mu \nu} R = T_{\mu \nu}$ can be
written as
\begin{eqnarray}
& & \dot{\rho} + (\dot{b} + \dot{\Lambda}) \rho = 0 
\; \; , \; \; \; 
\dot{\Lambda}^2 - \sum \dot{\lambda}_i^2 = 2 \rho 
\label{rho} \\
& & \ddot{\lambda}_i + \dot{\lambda}_i \dot{\Lambda} 
- \ddot{\Lambda} - \dot{\Lambda}^2 = - \rho + p_i 
= - (1 - w_i) \; \rho  \label{lambdai}
\end{eqnarray}
where overdots denote time derivatives. It follows from the
above equations that 
\begin{equation}\label{li} 
\ddot{\lambda}_i + \dot{\lambda}_i \dot{\Lambda} = C_i \; \rho
\; \; , \; \; \; \;
\ddot{\Lambda} + \dot{\Lambda}^2 = C_\Lambda \; \rho 
\; \; , \; \; \; \;
\ddot{b} + \dot{b} \dot{\Lambda} = C_b \; \rho 
\end{equation} 
where the constants $C_i$, $C_\Lambda = \sum C_i$ , and $C_b =
\sum w_i C_i$ are given by
\begin{equation}\label{cs}
C_i = w_i + \frac{1 - W}{D - 2} \; \; , \; \; \; 
C_\Lambda = \frac{D - 1 - W}{D - 2} \; \; , \; \; \; 
C_b = U + \frac{W - W^2}{D - 2} 
\end{equation}
with $W = \sum w_i$ and $U = \sum w_i^2$. The above equations
(\ref{rho}) -- (\ref{li}) have been solved recently by Chowdhury
and Mathur in \cite{cs}. See also \cite{b} for anisotropic
solutions in string/brane cosmology. We follow the methods of
\cite{cs} closely and define
\begin{equation}\label{hdefn}
H_i = \dot{\lambda}_i \; e^\Lambda \; \; , \; \; \; 
H = \dot{\Lambda} \; e^\Lambda \; \; , \; \; \;
B = \dot{b} \; e^\Lambda \; \; , \; \; \;
T = \rho \; e^\Lambda \; . 
\end{equation}
Note that $H = \sum H_i$, $B = \sum w _i H_i$, and that $T > 0$
since $\rho > 0$. It is straightforward to see that ($H_i, H, B,
T$) satisfy the equations
\begin{equation}\label{teqn}
\dot{H_i} = C_i \; T \; \; , \; \; \;  
\dot{H} = C_\Lambda \; T \; \; , \; \; \;  
\dot{B} = C_b \; T \; \; , \; \; \;  
\dot{T} = - \dot{b} \; T \; . 
\end{equation}
Define a parameter $\tau$ by $\dot{\tau} = T$, equivalently by
$t - t_0 = \int_0^\tau \frac{d \tau}{T}$. Then $\dot{f} = T
f_\tau$ for any function $f$ where a subscript $\tau$ denotes
$\tau$--derivative. Using the above definitions and equations,
one now gets
\begin{equation}\label{taueqn}
(\lambda_i)_\tau = \frac{H_i}{T \; e^\Lambda } \; \; , \; \; \;
\Lambda_\tau = \frac{H}{T \; e^\Lambda} \; \; , \; \; \;
b_\tau = - \frac{T_\tau}{T} = \frac{B}{T \; e^\Lambda} \; , 
\end{equation}
\begin{equation}\label{hsoln}
H_i = C_i \tau + K_i \; \; , \; \; \;  
H = C_\Lambda \tau + K_\Lambda \; \; , \; \; \;  
B = C_b \tau + K_b \; . 
\end{equation} 
It follows from equations (\ref{taueqn}) that 
$(T \; e^\Lambda)_\tau = H - B$ and, hence,
\begin{equation}\label{evsoln}
T \; e^\Lambda = \frac{C_\Lambda - C_b}{2} \; \tau^2 
+ (K_\Lambda - K_b) \; \tau + K_0 \; . 
\end{equation}
The constants $(t_0, K_i, K_\Lambda, K_b, K_0)$ in the above
equations are initial values of $(t, H_i, H, B, T \; e^\Lambda)$
respectively at $\tau = 0$. Equations (\ref{taueqn}), and then
equation $\dot{\tau} = T$, can now be solved to obtain
$(\lambda_i, \Lambda, b, T)$ and $t$ in terms of $\tau$. Thus
$(\lambda_i, \Lambda, b, T)$ are obtained implicitly in terms of
$t$.

\vspace{4ex}

{\bf 3. Asymptotic solutions and their properties}  

\vspace{2ex}

The general solutions to the above equations and a thorough
discussion of their properties are presented in detail in
\cite{cs}. Among other things it is shown that, for physical
ranges of parameters, the equation $T e^\lambda = 0$ has two
real roots $\tau_1$ and $\tau_2 > \tau_1$ and that the universe
expands into future with no singularities.

We are interested here in the asymptotic solutions in the limit
$t \gg 1$,\footnote{ Asymptotic solutions are valid when $\tau
\gg 1$ or $\tau_2 - \tau \ll \tau_2$ as the case may be, which
translates into $t \gg 1$ in natural units.}  which are also
given in \cite{cs}. In this limit, we have \footnote{ Here and
in the following $\simeq$ denotes that constant factors,
numerical as well as dimensionfull ones, which are not needed
here are omitted.}
\begin{equation}\label{asymp}
e^{\lambda_i} \simeq t^{\alpha_i} 
\; \; , \; \; \; 
e^{\Lambda} \simeq t^\alpha 
\; \; , \; \; \; 
e^b \simeq t^{\alpha_b}
\; \; , \; \; \; 
\rho \propto e^{- b - \Lambda} \simeq t^{- \alpha - \alpha_b}
\end{equation}
where $\alpha_i$'s are constants, $\alpha = \sum \alpha_i$, and
$\alpha_b = \sum w_i \alpha_i$. As shown in \cite{cs}, the limit
$t \gg 1$ corresponds to the following cases:

\vspace{2ex}

{\bf (i)} $C_\lambda > C_b$: 
In this case, $\tau \gg 1$ in the limit $t \gg 1$ and
\begin{equation}\label{alpha1}
\alpha_i = \frac{2 C_i}{C_\Lambda + C_b} \; \; , \; \; \;
\alpha = \frac{2 C_\Lambda}{C_\Lambda + C_b} \; \; , \; \; \;
\alpha_b = \frac{2 C_b}{C_\Lambda + C_b} \; . 
\end{equation}
Note that $\alpha + \alpha_b = 2$, and $\alpha > 1$ since
$C_\lambda > C_b$. 

\vspace{2ex}

{\bf (ii)} $C_\lambda < C_b$: 
In this case, $\tau \to \tau_2$ from below in the limit $t \gg 1$
and
\begin{equation}\label{alpha2}
\alpha_i = \frac{H_{i2}}{H_2} \; \; , \; \; \; 
\alpha = 1 \; \; , \; \; \;
\alpha_b = \frac{B_2}{H_2} 
\end{equation}
where $(H_{i2}, H_2, B_2)$ are the values of $(H_i, H, B)$
respectively at $\tau = \tau_2$. From the definitions of $H$ and
$B$ it follows that $H_2 = \sum H_{i2}$ and $B_2 = \sum w_i
H_{i2}$.

\vspace{2ex}

{\bf (iii)} $C_\lambda = C_b$: 
It is straightforward to obtain detailed solutions for this case
also, not given explicitly in \cite{cs}. However, it suffices
here to note the following.

For generic initial conditions $K_\Lambda \ne K_b$ and the
solutions depend on whether $K_\Lambda > K_b$ or $K_\Lambda <
K_b$. If $K_\Lambda > K_b$ then $\tau \gg 1$ in the limit $t \gg
1$ since $T e^\Lambda > 0$. Hence, this case can be thought of
as case {\bf (i)} with $C_\Lambda = C_b + \epsilon$ in the limit
$\epsilon \to 0_+$, equivalently with $\alpha \to 1$ from
above. The $\alpha$'s are then given by
\begin{equation}\label{alpha31}
\alpha_i = \frac{2 C_i}{2 C_\Lambda - \epsilon} \; \; , \; \; \;
\alpha = \frac{2 C_\Lambda}{2 C_\Lambda - \epsilon} 
\; \; , \; \; \;
\alpha_b = \frac{2 (C_\Lambda - \epsilon)}
{2 C_\Lambda - \epsilon} \; . 
\end{equation}

Similarly, if $K_\Lambda < K_b$ then $\tau \to \tau_3 \equiv
\frac{K_0}{K_b - K_\Lambda}$ from below in the limit $t \gg 1$.
Hence, this case can be thought of as case {\bf (ii)} with
$C_\Lambda = C_b - \epsilon$ in the limit $\epsilon \to
0_+$. The $\alpha$'s are then given by
\begin{equation}\label{alpha32}
\alpha_i = \frac{H_{i3}}{H_3} \; \; , \; \; \; 
\alpha = 1 \; \; , \; \; \;
\alpha_b = \frac{B_3}{H_3} 
\end{equation}
where $(H_{i3}, H_3, B_3)$ are the values of $(H_i, H, B)$
respectively at $\tau = \tau_3$. From the definitions of $H$ and
$B$ it follows that $H_3 = \sum H_{i3}$ and $B_3 = \sum w_i
H_{i3}$.

Note that $\alpha \ge 1$ for all the solutions. 

\vspace{4ex}

{\bf 4. Entropy, energy, and area}  

\vspace{2ex}

In the following, we assume that the spatial directions denoted
by $i = 1, 2, \cdots, n$ are compact and toroidal with
coordinate size $l_i$; the transverse spatial directions denoted
by $j = 1, 2, \cdots, m = D - 1 - n$ may be non compact, or
compact with coordinate sizes $l_j$ assumed to be $\gg l_i$. 


In a given direction with scale factor $e^\lambda$, the
coordinate and the physical sizes of the horizon within which
causal contact is possible are given respectively by
\begin{equation}\label{lh}
r_H = \int_0^t d t \; e^{- \lambda} \; \; , \; \; \; 
L_H = e^\lambda \; r_H \; \; . 
\end{equation}
Consider a compact direction, with coordinate size $l$ and scale
factor $e^\lambda$. Its physical size is then given by $L \simeq
e^\lambda l$. If its horizon size $r_H \ge l$, equivalently $L_H
\ge L$, then all of that direction is in causal contact and,
hence, $L_H$ is to be replaced by $L$. In the following, we take
this to be the case for $i = 1, 2, \cdots, n$; that is, we
consider the asymptotic limit $t > t_c \gg 1$ where $t_c$ is the
earliest time when $L_{H i} (t_c) > L_i$, $i = 1, 2, \cdots, n$.
Then $L_{H i}$ is to be replaced by $L_i$ and the coordinate and
the physical volume of the $n$ -- dimensional compact space,
within which causal contact is possible, are given respectively
by
\begin{equation}\label{vc}
v_c \simeq  \prod_{i = 1}^n l_i \; \; , \; \; \; 
V_c \simeq  \prod_{i = 1}^n L_i \; . 
\end{equation}
The coordinate and the physical volume of the horizon in the
transverse space are, respectively,
\begin{equation}\label{vnc}
v_{H \perp} \simeq \prod_{j = 1}^m r_{H j} \; \; , \; \; \;
V_{H \perp} \simeq  \prod_{j = 1}^m L_{H j} \; . 
\end{equation}
The total physical volume $V$ and the physical area $A$, within
which causal contact is possible, are thus given by
\begin{equation}\label{vph}
V \simeq V_c \; V_{H \perp} \; \; , \; \; \; 
A \simeq V_c \; (V_{H \perp})^{\frac{m - 1}{m}} \; \; . 
\end{equation}
When the transverse space is compact, let $t_\perp \gg t_c$ be
the earliest time when $L_{H j} (t_\perp) > L_j$, $j = 1, 2,
\cdots, m$. Then, equation (\ref{vph}) is still applicable for
$t \ll t_\perp$. For $t > t_\perp$, however, $L_{H j}$ is to be
replaced by $L_j$ and the coordinate and the physical volume of
the $m$ -- dimensional transverse space, within which causal
contact is possible, are given respectively by
\begin{equation}\label{vperp}
v_\perp \simeq  \prod_{j = 1}^m l_j \; \; , \; \; \; 
V_\perp \simeq  \prod_{j = 1}^m L_j \; . 
\end{equation}
The total physical volume $V$, within which causal contact is
possible, is now given by $V \simeq V_c V_\perp \simeq v_c
v_\perp e^\Lambda$.

It can be shown that the conservation equation $\dot{\rho} +
(\dot{b} + \dot{\Lambda}) \rho = 0$ implies that the comoving
entropy density $S_{co}$ is a constant, see the paper by
Tseytlin and Vafa in \cite{bv}. Following the standard method,
the physical entropy density $s$ is given by
\begin{equation}\label{s}
s = e^{- \Lambda} \; S_{co} \; . 
\end{equation}
The physical entropy $S$ and the physical energy $E$, defined
naturally as those contained in the volume $V$ within which
causal contact is possible, and assumed to be large as is
generically the case in an expanding universe, are then given by
\begin{equation}\label{sev}
S = s \; V
\; \; , \; \; \; 
E = \rho \; V 
\end{equation} 
using which $S$ may also be expressed in terms of $E$. 

We now evaluate the above quantities for the solutions in the
asymptotic limit $t \gg 1$, given in section {\bf 3}. $\;
e^{\lambda_i}$, $e^\Lambda$, and $\rho$ are then given by
equation (\ref{asymp}) from which it follows that $L_i \simeq
l_i \; t^{\alpha_i} \;$, $r_{H i} \simeq t^{1 - \alpha_i} \;$,
$L_{H i} \simeq t$, and
\begin{equation}\label{ab}
s \simeq t^{- \alpha} \; \; ,\; \; \; 
\rho \simeq t^{- \alpha - \alpha_b} \; . 
\end{equation}
We take $t > t_c \gg 1$ so that $L_{H i} > L_i$ for $i = 1, 2,
\cdots, n$ and, hence, are to be replaced by $L_i$ -- this is
possible if $\alpha_i < 1$ which is the case in the examples
below. Then, $V_c \simeq v_c \; t^{\alpha_c}$ where we define
\begin{equation}\label{alphacnc}
\alpha_c = \sum_{i = 1}^n \alpha_i \; \; , \; \; \;
\alpha_\perp = \alpha - \alpha_c 
= \sum_{j = 1}^m \alpha_j \; \; . 
\end{equation}
Note that, for increasing $t$, the compact space on the whole
expands if $\alpha_c > 0$ and contracts if $\alpha_c < 0$ since
$V_c$ increases or decreases respectively.

We now consider two cases where the transverse space is {\bf
(i)} non compact, and {\bf (ii)} compact.

\vspace{2ex}

\noindent
{\bf (i) Transverse space is non compact}  

\vspace{2ex}

We have $V_{Hnc} \simeq t^m$. The total physical volume $V$ and
the physical area $A$, within which causal contact is possible,
are then given by
\begin{equation}\label{avasymp}
V \simeq t^{m + \alpha_c} \; \; , \; \; \;
A \simeq t^{m - 1 + \alpha_c} \; \; .
\end{equation}
The physical entropy $S$ and the physical energy $E$ contained
in the volume within which causal contact is possible are given
by
\begin{equation}\label{seasymp}
S \simeq t^{m + \alpha_c - \alpha}
\; \; , \; \; \; 
E \simeq t^{m + \alpha_c - \alpha - \alpha_b} \; .
\end{equation}
Since $\alpha \ge 1$ for all the solutions, the entropy $S$
satisfies the Fischler -- Susskind holographic bound $S \le A$,
upto constant factors which are omitted here \cite{fsb}. The
entropy $S$ is given in terms of the energy $E$ as
\begin{equation}\label{se} 
S \simeq E^X \; \; , \; \; \; \; \; \; 
X = \frac{m + \alpha_c - \alpha} 
{m + \alpha_c - \alpha - \alpha_b} \; \; \; .  
\end{equation} 
The exponent $X$ then characterises the amount of entropy for
given energy: higher the value of $X$ higher is the
entropy. 

Note that for the $\alpha$'s obtained from equations
(\ref{alpha1}), $\alpha + \alpha_b = 2$ and therefore $X =
\frac{m + \alpha_c - \alpha} {m + \alpha_c - 2} \; \;$. Also
note that, generically, $\alpha$ and/or $\alpha_c$ and, hence,
the exponent $X$ depend on $n$, the dimension of the compact
space.

\vspace{2ex}

\noindent
{\bf (ii) Transverse space is compact}  

\vspace{2ex}

For $t_c < t \ll t_\perp$, the horizon size is less than the
size of the compact transverse space and, hence, the transverse
space is effectively non compact. The results in {\bf 4 (i)} are
then applicable here also.

For $t > t_\perp$, however, entire transverse space is in causal
contact and, hence, $L_{H j}$ are to be replaced by $L_j$, $j =
1, 2, \cdots, m$, and $V_{H \perp}$ is to be replaced by
$V_\perp \simeq t^{\alpha_\perp}$. Then, the total physical
volume $V \simeq t^\alpha$. \footnote{The area is now zero since
a compact space has no boundary. Therefore, more rigorous
Bousso's formulation \cite{fsb} has to be used to verify the
holographic principle, which will not be pursued here.} The
physical entropy $S$ and the physical energy $E$ contained in
the volume within which causal contact is possible are then
given by
\begin{equation}\label{seasympii} 
S \simeq v_c v_\perp S_{co} \simeq constant \; \; , \; \; \; 
E \simeq t^{- \alpha_b} \; .  
\end{equation} 

The entropy $S$ is constant because the comoving entropy density
$S_{co}$ is constant \cite{bv} and, since $t > t_\perp$, the
comoving coordinate volume of the region within which causal
contact is possible is now that of the entire space, namely $v_c
v_\perp$, which is also constant; the entropy $S$ which is a
product of these two quantities is, therefore, constant.

The energy varies -- typically decreases -- because the physical
size of the space varies -- typically expands -- and work is
done against/by the pressure. Thus if in all directions either
there is no expansion/contraction or there is no pressure then
no work is done and, hence, the energy $E$ must be constant.
This is indeed the case since either no expansion/contraction or
no pressure in $i^{th}$ direction means either $\alpha_i = 0$ or
$w_i = 0$ and thus $w_i \alpha_i = 0$; when this is the case for
all directions it follows that $\alpha_b = \sum w_i \alpha_i =
0$. The energy $E \simeq t^{- \alpha_b}$ is then constant.

Let $\alpha_b = 0$. Then, for $t > t_\perp$, $S$ and $E$ are
both constants and, a priori, the relation between them can be
anything. However, for $t_c < t \ll t_\perp$, equation
(\ref{se}) is applicable from which we see that $X = 1$ since
$\alpha_b = 0$ and, hence, $S \simeq E$. If one assumes that
this relation holds even for $t > t_\perp$, as seems physically
reasonable, then it follows that $S \simeq E \simeq constant \;$
for $t > t_\perp$ also.

In the following, unless mentioned otherwise, we assume that the
transverse space is non compact; or that if the transverse space
is compact then we consider only $t \ll t_\perp$ so that the
transverse space is effectively non compact. The results for $t
> t_\perp$ are straightforward to obtain.

\vspace{4ex}

{\bf 5. Examples}  

\vspace{2ex}

{\bf (i)} 
Consider an example where $\lambda_i = 0$ for the compact
directions $i = 1, 2, \cdots, n$. Then $\dot{\lambda}_i = 0$
and, as follows from equations (\ref{lambdai}) and (\ref{li}),
$w_i = \sigma$ and $C_i = 0$ for $i = 1, 2, \cdots, n$ where
$\sigma$ is yet to be determined. Thus, the compact space needs
to be supported by an isotropic pressure $p_c = \sigma \rho$ in
order for its scale factors to remain constant.

Define $W_\perp = \sum_\perp w_j$ and $U_\perp = \sum_\perp
w_j^2$ where the sum in $\sum_\perp$ is over the transverse
directions $j = 1, \cdots, m$ only. It then follows, after some
algebra, that $\sigma = \frac{W_\perp - 1}{m - 1} \; $ and
\begin{eqnarray}
C_j = w_j + \frac{1 - W_\perp}{m - 1} 
& , & j = 1, \cdots, m \\
C_\Lambda = \frac{m - W_\perp}{m - 1} & , &
C_b = U_\perp + \frac{W_\perp - W_\perp^2}{m - 1} 
\; \; . \label{ci}
\end{eqnarray}
These expressions for the $C$'s and, hence, the subsequent
analysis and the results are identical to those of an $(m + 1)$
-- dimensional universe. In particular, they are independent of
$n$, the dimension of the compact space which is now kept at a
constant size by a pressure $p_c = \sigma \rho$.

Furthermore let $w_j = \omega$ for $j = 1, \cdots, m$. Thus, the
transverse space is isotropic and contains a perfect fluid with
pressure $p_\perp = \omega \rho$. Then $W_\perp = m \omega \;$,
$\sigma = \frac{m \omega - 1} {m - 1} \;$, $C_i = 0$ for $i = 1,
2, \cdots, n$ by construction, and
\begin{eqnarray}
C_j = \frac{1 - \omega}{m - 1} 
& , & j = 1, \cdots, m \\
C_\Lambda = \frac{m \; (1 - \omega)}{m - 1} & , & 
C_b = \frac{m \; \omega \; (1 - \omega)}{m - 1} 
\; \; . \label{ciso} 
\end{eqnarray}
Note that $C_\Lambda \ge C_b$ since $\omega \le 1$. Also, it is
physically reasonable to think of $\omega = 1$ as $\omega \to 1$
from below, equivalently as $C_\Lambda \to C_b$ from above.
Hence, $(\alpha_i, \alpha, \alpha_b)$ are given by equations
(\ref{alpha1}), and $(\alpha_c, \alpha_\perp)$ by equations
(\ref{alphacnc}). Thus, $\alpha_i = \alpha_c = 0$ for $i = 1, 2,
\cdots, n$, $\alpha_\perp = \alpha$, and $(\alpha_j, \alpha,
\alpha_b)$ for $j = 1, \cdots, m$ are given by
\begin{equation}\label{cnc}
\alpha_j = \frac{2}{m \; (1 + \omega)} \; \; , \; \; \;
\alpha = \frac{2}{1 + \omega} \; \; , \; \; \;
\alpha_b = \frac{2 \; \omega}{1 + \omega} \; . 
\end{equation}
The entropy $S$ is given in terms of the energy $E$ as $S \simeq
E^{X_\omega(m)}$ where, since $\alpha_c = 0$,
\begin{equation}\label{x1}
X_\omega(m) = \frac{m - \alpha} {m - 2} \; \; , \; \; \; 
\alpha = \frac{2}{1 + \omega} \; \; .
\end{equation}
Note that $X_\omega$ and, hence, the entropy $S$ are independent
of $n$, the dimension of the compact space and are identical to
those of an $(m + 1)$ -- dimensional universe. Also, note that
the exponent $X_\omega$ is given by $X_0(m) = 1$ for $\omega =
0$, and by $X_1(m) = \frac{m - 1} {m - 2}$ for $\omega = 1$.

\vspace{2ex}

{\bf (ii)} 
Consider a second example where $w_i = \sigma$ for $i = 1, 2,
\cdots, n$ with $n \ge 1$ and $w_i = \omega$ for $i = n + 1,
\cdots, D - 1$. Hence, $\alpha_1 = \alpha_2 = \cdots =
\alpha_n$, $\alpha_{n + 1} = \cdots = \alpha_{D - 1}$, $\alpha_c
= n \alpha_1$, $\alpha_\perp = m \alpha_{n + 1}$, and $\alpha =
\alpha_c + \alpha_\perp$. We assume that $C_\Lambda > C_b$ or $
\to C_b$ from above. Then, $(\alpha_i, \alpha, \alpha_b)$ are
given by equations (\ref{alpha1}) and $\alpha + \alpha_b =
2$. Using equations (\ref{cs}), it follows that $C_b - C_\Lambda
= \frac{(1 - \omega)^2} {D - 2} \; f(x)$ where $x = \frac{1 -
\sigma}{1 - \omega}$ and
\begin{equation}\label{f}
f(x) = n (m - 1) x^2 - 2 n m x + m (n - 1) \; \; . 
\end{equation}
Hence, $C_\Lambda \ge C_b$ if $x_- \le x \le x_+$ where 
\begin{equation}\label{xpm}
x_\pm = \frac{n m \pm \sqrt{\Delta}}{n \; (m - 1)} 
\; \; , \; \; \; 
\Delta = n m (n + m - 1) 
\end{equation}
are the zeroes of $f$. Note that, for the example {\bf (i)}, 
$x = \frac{m}{m - 1}$ which is where $f(x)$ is minimum. It is
easy to show that $\alpha_i = \frac{\alpha_c}{n} \ge 0$ if $x
\le \frac{m}{m - 1}$ and $ < 0$ if $x > \frac{m} {m - 1}$
for the compact directions $i = 1, 2, \cdots, n$.

Consider the limit $\alpha \to 1$ from above. Then, $C_\Lambda
\to C_b$ from above and $x \to x_\pm$ from within its allowed
range. At $x = x_\pm$, we have $\alpha = \alpha_b = 1$ and,
after some algebra, 
\[
\alpha_{c \pm} = \frac{n \mp \sqrt{\Delta}}{n + m} \; \; .
\]
$\alpha_i$'s then follow from $\alpha_c = n \alpha_1$ and
$\alpha_\perp = \alpha - \alpha_c = m \alpha_{n + 1}$. The
entropy $S$ is given in terms of the energy $E$ as $S \simeq
E^{X_\pm}$ where, since $\alpha = 1$,
\begin{equation}\label{x2}
X_\pm = \frac{m + \alpha_{c \pm} - 1} 
{m + \alpha_{c \pm} - 2} \; \; . 
\end{equation}
Note that $\alpha_{c \pm}$ and, hence, $X_\pm$ and the entropy
$S$ now depend also on $n$, the dimension of the compact space.
Also note that $X_- < X_1(m)$ since $\alpha_{c -} > 0$, and $X_+
> X_1(m)$ since $\alpha_{c +} < 0$ where $X_1(m) = \frac{m -
1}{m - 2}$ is defined in equation (\ref{x1}). Some examples of
$(n, m, X_+)$ are:
\begin{equation}\label{nmx}
(2, 8, \frac{6}{5}) \; \; , \; \; \; 
(5, 5, \frac{3}{2}) \; \; , \; \; \; 
(3, 6, \frac{4}{3}) \; \; , \; \; \; 
(7, 3,  \sim 4.07) \; \; . 
\end{equation}

One would expect the most entropic object in an $(m + 1)$ --
dimensional spacetime to be a Schwarzschild black hole or an
isotropic universe containing matter with $\omega = 1$. For both
of them, the exponent $X$ in the relation $S \simeq E^X$ is
given by $X = X_1(m)$. Therefore, the fact that $X$ can be $>
X_1(m)$ is perhaps surprising. Here, we have an $m$ --
dimensional transverse space and an $n$ -- dimensional compact
space. As pointed out before, the entropy satisfies the
holographic bound, so its violation can not be the reason for
$X$ being $> X_1(m)$. The origin of this behaviour of $X$ can,
however, be traced back to the fact that $\alpha_c < 0$; hence,
in the limit $t \gg 1$, the compact space is contracting to zero
size; equation (\ref{seasymp}) then leads to a value of $X_+$
which is $> X_1(m)$. See near the end of next section for more
discussion.

\vspace{4ex}

{\bf 6. Fractional branes in the universe}  

\vspace{2ex}

We assume that our observed universe is described by string/M
theory. At low temperatures, the evolution of the universe is
described in the standard way by a low energy effective action.
At early times, when the temperature is of the order of string
scale, higher modes of the strings are excited and the evolution
must be described by stringy variables \cite{bowick,k1}.

Assuming that our universe originated from highly excited and
highly interacting strings, we have proposed in \cite{k2} a
maximum entropic principle to determine the number $(3 + 1)$ of
large spacetime dimensions. According to this principle, the
spacetime configuration that eventually emerges from the highly
excited and interacting strings is the one that has maximum
entropy for a given amount of energy, both assumed to be
large. In \cite{k2}, we considered an isotropic universe with a
compact space of constant size which is assumed to be
$\stackrel{>} {_\sim} l_s$, the string length. Hence, the
entropy $S \simeq E^{X_\omega(m)}$ as given in equation
(\ref{x1}). The maximum entropic principle then leads to a $(3 +
1)$ -- dimensional spacetime. As can be easily checked, this
result remains valid also for the anisotropic examples
considered here although, contrary to an assumption in
\cite{k2}, the compact space contracts to zero size in some
cases.

The evolution of the universe at early times, when its
temperature is high, is not well understood. Within the context
of perturbative string theory, a natural idea has been to assume
that the universe consists of gas of strings, their winding
modes, and/or gases of various branes, and then obtain the
evolution of the universe using a low energy effective action
\cite{bv, branes}. In some of these models, the resulting
cosmology is anisotropic and the corresponding anisiotropic
solutions have also been obtained \cite{b}.

Recently, Chowdhury and Mathur proposed in \cite{cs} a novel
model where the early universe consists of mutually BPS
intersecting brane and antibrane configurations. In such BPS
configurations, the intersecting branes (and similarly
antibranes) form bound states and become fractional, supporting
very low energy excitations and creating thereby large entropy
for a given energy. Hence, the early universe is likely to
contain, and be dominated by, such fractional brane
configurations because of their high entropy. These
configurations, as Chowdhury and Mathur explain clearly, are
different from string/brane gas in \cite{bv, branes, b}.

Assuming all the spatial directions to be toroidal, and the
brane antibrane decay or annihilation to be negligible,
Chowdhury and Mathur also obtain the energy momentum tensor,
$T^\mu \; _\nu = diag \; (- \rho, \; p_i)$ where $p_i = w_i
\rho$, for these configurations with net brane charges
vanishing. They also solve in complete generality the Einstein's
equations of motion with arbitrary $w_i$'s, and discuss the
properties of the solutions. See \cite{cs} for details; see
\cite{b} also.

An $N$ charge configuration of the fractional branes consists of
$N$ or $N - 1$ stacks, each containg a large number $n_I$ of
coincident branes and an equal number of antibranes,
intersecting in a mutually BPS way; in the later case, the $N -
1$ stacks also have a common boost with $n_I$ units of Kaluza --
Klein momentum. See \cite{cs} for intersection rules. For an $N$
charge configuration, the energy $E$ and the entropy $S$ are
given by
\begin{equation}\label{nes}
E = 2 \sum_{I = 1}^N n_I m_I  \; \; , \; \; \; 
S \simeq \prod_{I = 1}^N \sqrt{n_I} 
\end{equation} 
where $m_I$ is the mass of the $I^{th}$ type of brane/boost and
is constant; $m_I = \tau_I V_I$ for branes and $= \frac{1}{R}$
for boosts where $\tau_I$ and $V_I$ are the tension and volume
of the brane and $R$ is the size of the compact boost direction.
The numbers $n_I$, in thermal equilibrium, are obtained by
maximising the entropy $S$ with respect to $n_I$ keeping the
energy $E$ fixed. It then follows that $n_I m_I = \frac{E}{2 N}$
and, hence, that $S \simeq E^\frac{N}{2}$. 

This relation is derived with the volume of the system kept
fixed, see \cite{cs} and the references therein for details. As
pointed out by a referee, the entropy $S$ and the energy $E$
defined in section {\bf 4} in an expanding universe will, in
general, have a different relation since now the volume changes
as dictated by the relevant equations of motion. \footnote{ For
example, for radiation in a $m + 1$ -- dimensional spacetime, $p
= \frac{\rho}{m}$ and, with volume kept fixed, $S \simeq
E^{\frac{m}{m + 1}}$. However, in an isotropic expanding
universe the volume changes and $S$ and $E$ defined in section
{\bf 4} obey a different relation $S \simeq E^{\frac{m (m -
1)}{(m + 1) (m - 2)}}$.} However, note that the $N = 3, 4$
configurations considered here, when in a finite volume,
describe black holes. The universe containing such
configurations may therefore be thought of as containing black
hole fluid in a $m + 1$ -- dimensional non compact spacetime.
Such a fluid is known to obey the equation of state $p = \rho$
and the corresponding $S$ and $E$ defined in section {\bf 4}
then obey the relation $S \simeq E^{\frac{m - 1}{m - 2}}$
\cite{bf}. Thus, the entropy $S$ is expected to be $\simeq
E^\frac{3}{2}$ for the 3 charge case and $\simeq E^2$ for the 4
charge case. Hence, for these cases, the expected form turns out
to be $S \simeq E^{\frac{N}{2}}$, the same as that obtained with
volume kept fixed which, we reemphasise, is not the case in
general. The 4 charge case may thus provide a detailed
realisation of the maximum entropic principle proposed in
\cite{k2}.

Let the $N$ charge fractional branes be wrapped along the $n$ --
compact directions and smeared uniformly in the transverse
space. We assume that the transverse space is non compact; or
that if it is compact then its size is sufficiently large so as
to correspond to our observed universe. The $w_i$'s in the
equation of state $p_i = w_i \rho$ for fractional branes may be
obtained as in \cite{cs}. We assume that the evolution of the
fractional branes in the universe is given by the evolution of
the anisotropic universe with the corresponding $w_i$'s; and
that the physical quantities are similarly related. \footnote{
The process of intersecting branes forming bound states,
becoming fractional, and creating large entropy is a non local
process extending over the brane volume, namely over the size of
the compact directions $i = 1, 2, \cdots, n$ along which the
branes wrap. In the asymptotic limit $t > t_c >> 1$ considered
here, there is sufficient time for this process to reach thermal
equilibrium and, hence, the $w_i$'s for the anisotropic universe
may be assumed to be those of the fractional branes obtained as
in \cite{cs}.} Hence, for example, the entropy of the fractional
branes in the universe is assumed to be given by the entropy of
the anisotropic universe with the corresponding $w_i$'s, which
can be calculated using the results presented here. We now
compare the entropy $S$ of the fractional branes in the
universe, thus obtained, with its expected value $\simeq
E^\frac{N}{2}$.

We consider $3$ and $4$ charge configurations here and denote
by, for example, 25B a $3$ charge configuration consisting of
two intersecting stacks of M2 and M5 branes and with a boost
along the common direction; similary, by 2255 a $4$ charge
configuration consisting of two stacks of M2 branes and two
stacks of M5 branes intersecting. Note that a given $N$ charge
configuration can be described equivalently in several ways, all
related by a series of S, T, and U dualities. Thus, the $3$
charge configurations 222 and 25B are equivalent; and,
similarly, the $4$ charge configurations 2255 and 555B.

The $w_i$'s for these configurations, calculated as in
\cite{cs}, turn out to be of the form $w_i = \sigma$ for the
compact directions $i = 1, 2, \cdots, n$ and $w_i = 0$ for the
transverse directions $i = n + 1, \cdots, D - 1$. Hence,
$\alpha_1 = \alpha_2 = \cdots = \alpha_n \;$, 
$\alpha_{n + 1} = \cdots = \alpha_{D - 1} \;$, 
$\alpha_c = n \alpha_1 $, 
$\alpha_\perp = m \alpha_{n + 1} \;$, 
$\alpha = \alpha_c + \alpha_\perp \;$, and 
$\alpha_b = n \sigma \alpha_1$ where $n + m = D - 1 \;$
with $D = 11$. 

Using the formulas in \cite{cs}, it follows that $(m, \sigma) =
\left( 4, - \frac{1}{3} \right)$ for 3 charge configurations,
$(m, \sigma) = \left( 3, - \frac{1}{2} \right)$ for 4 charge
configurations, and $n = D - 1 - m$. It can be seen that
$C_\Lambda > C_b$ for these configurations and, hence,
$(\alpha_i, \alpha, \alpha_b)$ are given by equations
(\ref{alpha1}). Using these equations one gets, for all these
configurations,
\begin{equation}\label{34}
\left( \alpha_1, \alpha_c, \alpha_{n + 1}, \alpha, \alpha_b
\right) = \left( 0, 0, \frac{2}{m}, 2, 0 \right) \; .
\end{equation} 

Note that these configurations are a particular case of example
{\bf (i)}, now with $\omega = 0$ and, accordingly, $\sigma = -
\frac{1}{m - 1}$, $\alpha_i = 0$ for $i = 1, 2, \cdots, n$ and
$(\alpha_{n + 1}, \alpha, \alpha_b)$ as given in equation
(\ref{cnc}). Also, note that either $\alpha_i = 0$ or $w_i = 0$
for all directions and, hence, $\alpha_b = \sum w_i \alpha_i =
0$. It then follows from equation (\ref{se}) that $X =
1$. Hence, the entropy $S$ for all of these 3 and 4 charge
configurations is given by $S \simeq E$. 

The entropy is not of the form expected from considerations
given in the paragraph containing footnote 5, namely $\simeq
E^\frac{3}{2}$ for 3 charge cases and $\simeq E^2$ for 4 charge
cases. Given the importance of 4 charge fractional branes in
understanding the early universe, it is important to study how
to obtain the expected form for $S$. Hence, we now consider a
few possibile ways of achieving this.

{\bf (i) } 
The entropy $S$ and the energy $E$ above are those contained in
the volume within which causal contact is possible. It may be
that the non local fractionation process extends not only over
the brane volume, namely over the compact space, but over the
entire transverse space also. Then it is the total entropy
$S_{tot}$ and total energy $E_{tot}$ contained in the entire
transverse space which should obey the expected relation.
\footnote{ This possibility is pointed out by a referee.} If the
transverse space is non compact then $S_{tot}$ and $E_{tot}$ are
infinite. Hence, consider a large but compact transverse
space. Then $S_{tot}$ and $E_{tot}$ are given by $S$ and $E$
when $t > t_\perp$ and are constants for the present cases since
$\alpha_b = 0$. However, as discussed in section {\bf 4 (ii)},
the relation between these constant quantities is likely to be
$S_{tot} \simeq E_{tot}$ and not $S_{tot} \simeq
E_{tot}^{\frac{N}{2}}$.

{\bf (ii)} 
One may define a `holographic' entropy $S_{hol}$ as the maximum
possible entropy in the physical volume $V$. In the case of non
compact transverse space, or for $t \ll t_\perp$ in the case of
compact transverse space, $S_{hol} \simeq A$, the area bounding
the volume $V$. From equations (\ref{avasymp}) and
(\ref{seasymp}), it follows that $S_{hol}$ is given in terms of
the energy $E$ as
\begin{equation}\label{sehol} 
S_{hol} \simeq E^{X_{hol}} \; \; , \; \; \; \; \; \; 
X_{hol} = \frac{m + \alpha_c - 1} 
{m + \alpha_c - \alpha - \alpha_b} \; \; \; .  
\end{equation} 
Note that for any $w_i$s for which $\alpha_c = 0$ and $\alpha +
\alpha_b = 2$, we have $X_{hol} = X_1(m) = {\frac{m - 1} {m -
2}}$. Applied to 3 and 4 charge cases, one obtains the expected
relation $S_{hol} \simeq E^{\frac{N}{2}}$. This relation is also
likely to hold between the corresponding total $S_{hol}$ and
$E_{tot}$, defined as in the possibilty {\bf (i)} above. Hence,
may be it is the holographic entropy $S_{hol}$ which should obey
the expected relation.

{\bf (iii)}
Another method of obtaining the expected relation may be the
following. Let us view the 3 and 4 charge configurations as a
particular case of example {\bf (i)} with $\omega = 0$. The
entropy $S \simeq E$ obtained for the brane configurations above
are then a particular case of equation (\ref{x1}) with $\omega =
0$.

Seen as a particular case of example {\bf (i)}, it is now clear
how to obtain the expected entropy for the fractional brane
configurations: one assumes that $\sigma = \frac{m \omega - 1}{m
- 1}$, as in example {\bf (i)}, and that $\omega \to 1$. Then
the entropy $S$ is given by equation (\ref{x1}) with $\omega \to
1$, that is $S \simeq E^{X_1(m)}$ with $X_1(m) = {\frac{m - 1}
{m - 2}}$. Applied to 3 and 4 charge cases, one obtains the
expected relation $S \simeq E^{\frac{N}{2}}$.

The possibility {\bf (iii)} may be interpreted physically in the
following way.  In \cite{cs}, the $w_i$'s are obtained assuming
that the fractional brane and antibrane configurations, which
wrap the compact directions, are not decaying or annihilating
(or they do so very slowly and negligibly). Thus, the compact
space has negative pressure, produced by the tension of the
branes wrapping it, and the transverse space has no decay
products.  Then, as the caluclations in \cite{cs} show, one gets
$w_i = - \frac{1}{m - 1}$ for compact directions and $w_i = 0$
for transverse directions.

However, the branes and antibranes are expected to eventually
decay and annihilate each other and emit radiation, massless
scalars, etcetera as decay products. Thus, the compact
directions will eventually be relieved of their negative
pressure because the branes wrapping them decay and annihilate,
and the space will be full of decay products. In particular,
this implies that if the brane antibrane dynamics are fully
taken into taken into account then the $w_i$'s will be different
from the ones used above. It is physically reasonable that
massless scalars dominate the decay channels since they are
generically present in string theory and are entropically
favourable. It is then likely that $w_i = \omega \to 1$ for the
transverse directions.

One also has to understand why $w_i = \sigma$ for the compact
directions $i = 1, 2, \cdots, n$ and why $\sigma$ is related to
$\omega$ as in the example {\bf (i)}. For this purpose, note
that the 3 and 4 charge configurations of M theory considered
here can be described equivalently by those of string theory
ones, namely D1D5B and 3333 respectively, obtained by a series
of S, T, and U dualities. This means that the physical
quantities should be the same for these equivalent
descriptions. Using the results of \cite{cs}, which are
applicable for these cases also, it is easy to show that $m$,
$\sigma$, and $\alpha$'s remain the same as before but, for
string theory, $n = 10 - m$ since $D = 10$.

Considering 222 and 3333 configurations, it is clear that the
compact space may be taken to be isotropic and, hence, $w_i =
\sigma$ for the compact directions. But, it is not clear why or
how the relation $\sigma = \frac{m \omega - 1}{m - 1}$ is
enforced. This relation follows, as in the example {\bf (i)}, if
one assumes that the sizes of the compact directions remain
constant. But the physical reason for this assumption is not
immediately clear.

Now, note that example {\bf (ii)} is applicable here even if
$\sigma \ne \frac{m \omega - 1}{m - 1}$. From the results
presented here, it is clear in particular that the exponent $X$
that characterises the amount of entropy depends in general on
$n$, the dimension of the compact space. The only exception is
the case of example {\bf (i)}.

Since the 3 and 4 charge configurations considered here can be
described equivalently by M theory or by string theory, it
follows that the physical quantities, in particular the entropy
$S$, should not depend on $n$, the dimension of the compact
space since the $n = 10 - m$ in string theory and the $n = 11 -
m$ in M theory are different. This independence is possible if
the fractional branes obey the relation $\sigma = \frac{m \omega
- 1}{m - 1}$ throughout their evolution in the universe,
including their decay and annihilation. Since this follows from
the requirements of S, T, U duality symmetries, it is likely
that the relation between $\sigma$ and $\omega$ is enforced by
these same symmetries.

The T duality symmetries of string theory also have another
implication. In example {\bf (ii)}, $\alpha_c < 0$ if $x \ge
\frac{m}{m - 1}$. The compact space then contracts to zero size.
For toroidal compactifications, and perhaps more generally also,
the T duality symmetry ensures that the physical size of the
compact space remains $\stackrel{>}{_\sim} l_s$, the string
length. This implies that T duality symmetry must also ensure
that the physical $w_i$'s will be such that $\alpha_i \ge 0$ for
the compact directions.

In the example {\bf (ii)} then $x$ must be $ \le \frac{m} {m -
1}$ by T duality symmetry. Then $\alpha_c$ remains $\ge 0$;
hence, the exponent $X$ which characterises the amount of
entropy remains $\le X_1(m) = \frac{m - 1}{m - 2}$, in accord
with one's expectation that the most entropic object in an $(m +
1)$ -- dimensional spacetime is a Schwarzschild black hole or an
isotropic universe containing matter with $\omega = 1$.

\vspace{4ex}

{\bf 7. Conclusion}  

\vspace{2ex}

We conclude by mentioning a few issues for further study. In
this paper, we considered only two special sets of $w_i$'s which
were sufficient for our purposes here. It is desireable to
consider a general set of $w_i$'s, in the range $- 1 \le w_i \le
1$, and obtain the general conditions on $w_i$'s required to
realise the expected entropy for the 3 and 4 charge
configurations; the conditions required to ensure that their
entropy, equivalently the exponent $X$ in equation (\ref{se}),
remains the same in the equivalent string and M theoretic
descriptions; and also the conditions required to ensure that
the compact sizes remain $\stackrel{>}{_\sim} l_s$ as required
by T duality symmetry.

In this context, it may also be useful to study the implications
of S, T, U duality symmetries at the level of equations of
motion and their solutions.

As we have seen here, the expected entropy for the 3 and 4
charge fractional branes can be realised under some conditions
which are likely to be enforced by S, T, U dualities of the
string/M theory. Then the 4 charge configurations, having the
highest entropy for a given energy, may indeed provide a
detailed realisation of the maximum entropic principle proposed
in \cite{k2} to determine the number $(3 + 1)$ of large
spacetime dimensions. It is therefore important to study in
detail the decay and annihilation processes in the 4 charge
fractional brane configurations in the universe, and to
understand the role of S, T, U duality symmetries of string/M
theory in this setting. 

It is also important to explore other ways of obtaining the
expected form for entropy, besides those considered here.

\vspace{3ex}

{\bf Acknowledgement:} 
We thank S. D. Mathur for a private correspondence and the
referees for pointing out \cite{b}, for their comments, and for
helpful suggestions which improved the paper.


\end{document}